\documentclass[12pt]{article}

\author{Vladimir A. Petrov \footnote{e-mail:Vladimir.Petrov@ihep.ru}}

\title{On the momentum transfer dependence of the phase in elastic scattering.}
\date{}

\begin{document}

\maketitle
A. A. Logunov Institute for High Energy Physics, NRC KI,
Protvino, RF

\begin{abstract}

The question is discussed: to what extent the often assumed independence of the phase of the elastic scattering amplitude from the momentum transfer \textit{in the region of only small values} of $ t $ limits $ t $-dependence of the phase generally.
Analyticity allows to give  a proof that if the phase of a strong interaction scattering amplitude is independent on the transferred momentum $ t $  at small values of $ t $ then it does so in all physical region. Moreover, if such an independence holds in any physical energy $ \sqrt{s} $ region including the values infinitesimally close to the first inelastic threshold from below the whole scattering amplitude vanishes. Relationship of the $ t $-dependence of the phase with the size of the interaction range is also discussed.
\end{abstract}

\section{Introduction}	

It was mentioned more than once (see, e.g.\cite{Ku} ) that the use of the popular Bethe parametrization, suggested as early as in 1958 \cite{Be} , for the scattering amplitude with account of Coulomb exchanges
\begin{equation}
T_{C+N} (s;t) = T_{C} e^{i\alpha\varphi(s,t)}+ T_{N} (s;t)
\end{equation}
where
\begin{center}
$T_{C} = 8\pi s \alpha F^{2}(t)/t  $
\end{center}
(with $F(t)$ the electric form factor)
leads to the phase $ \Phi_{N}(s;t) $ of the strong interaction amplitude\footnote{$ T_{N}(s;t)\equiv T_{N} (s+i0, t), s \geq s_{el} $ .}  $ T_{N}(s;t)=\mid T_{N}(s;t)\mid \exp (i\Phi_{N}(s;t))$ which is independent on $ t $.

In phenomenological practice  this independence makes many things much easier and is often assumed (see e.g.\cite{We} ) and used to extract the ratio
\begin{center}
$ \rho (s)\doteq Re T_{N}(s;0) / Im T_{N} (s;0)$
\end{center}
from the experimental data.

\section{Analyticity}	
In many cases, however, such a $ t $-independence is supposed to hold "for small values of $ t $" only.
Below we consider to which extent this reservation allows " freedom of choice" at arbitrary transfers.  

To formalize the assumption in question, let us consider the quantity
  \begin{center}
 $ \rho (s,t) = Re T_{N}(s;t) / Im T_{N} (s;t) = \cot \Phi_{N}(s,t) $
 \end{center}
where $  T_{N}(s;t) $ is taken for simplicity the elastic scattering amplitude of identical strongly interacting scalars with $ s_{el}= 4m^{2} $. 

Both $ Re T_{N}(s;t) $ and $ Im T_{N} (s;t) $ which are real and imaginary parts of the boundary value $ T_{N}(s+i0,t) $ of an analytic function $ T_{N}(s,t) $ at physical $ s $ are analytic in $ t $ in the Martin ellipse \cite{Ma} 
\[\mid\ 4p^{2}+t\mid +\mid t\mid < 4p^{2}(1 + 4m^{2}/p^{2})^{1/2}\]
with focii at $ t=0 $ and $ t=-4p^{2} = s+4m^{2} $ and semi-major axis $ 2p^{2}(1+ 4m^{2}/p^{2})^{1/2} $.

Assume that $ \rho (s,t) $  ( and thus $ \Phi_{N}(s;t) $) is  $ t $-independent in the  interval $ (-\tau \leq t \leq 0] $
where $ \tau > 0 $ is arbitrarily small while $ s $ takes physical values in some interval on $ [4m^{2}, +\infty) $.
The following theorem then holds.

\textbf{Theorem I.}

\textit{ Let  $ \rho (s;t)$ is independent on $ t $ at $ t\in (-\tau, 0] $ and at  $ s \in [s_{1},s_{2}] \subset [4m^{2}, +\infty)  $. Then  $  \rho (s;t) $ is independent on $ t $ inside the Martin ellipse and, in particular, in the whole physical region
$ -4p^{2} \leq t \leq 0 $ with $ s \in \forall [s_{1},s_{2}] \subset [4m^{2}, +\infty)$.}

Proof.
The assumed $ t  $ - independence of $ \rho (s;t)\equiv \rho (s) $  at $ -\tau < t \leq 0 $ and physical $ s \in \forall [s_{1},s_{2}] \subset [4m^{2}, +\infty) $ implies the following relationship 
\begin{equation}
Re T_{N}(s;t)\mid_{[-\tau \leq t \leq 0]}  =  \rho (s)\cdot Im T_{N} (s;t)\mid_{[-\tau \leq t \leq 0]}.
\end{equation}
The function 
\[f(t)\doteq Re T_{N}(s;t) - \rho (s)\cdot Im T_{N} (s;t)\]
at fixed $ s $ as defined in the hypothesis of the theorem 

1) is analytic inside the Martin ellipse and

2) vanishes on the interval $ (-\tau \leq t \leq 0] $ lying inside the analyticity 

domain.

From the uniqueness theorem for analytic functions \cite{Pri} we conclude that
\[f(t)\equiv 0,\]
inside the Martin ellipse and $ s \in \forall [s_{1},s_{2}] \subset [4m^{2}, +\infty) $. This holds, in particular, at all physical values of $ t $ which are contained in this ellipse. So, $ \rho (s;t)\equiv \rho (s)  $, and hence the phase $ \Phi_{N}(s;t)$, does not depend on $ t $ at all physical $ t $:
\[\Phi_{N}(s;t) = \arctan (1/\rho(s))\] Q.E.D.

 Note that for the proof we could use even the narrower analyticity domain, the Lehmann ellipse.

\section{Unitarity}

The unitarity condition enables us to prove the following 

\textbf{Theorem II}

\textit{If the phase $\Phi_{N}(s;t) $ of the physical scattering amplitude 
\[T_{N}(s;t)=\mid T_{N}(s;t) \mid\exp (i\Phi_{N}(s;t))\] is independent on $ t $ at $ t\in (-\tau, 0] $, where a positive number $ \tau  $ is arbitrarily small and the energy region contains the values $s \in [s_{1},s_{2}]\subset [s_{inel}-\Delta s,s_{inel})$,where $ s_{inel}$ denotes the lowest inelastic threshold and $\Delta s \ll s_{inel} $, then $ T_{N}(s,t) \equiv 0 $ in all physical region of $ t $ and $ s $.}

Proof.
From the Theorem I  proved in the previous Section we deal with the amplitude 
$ T_{N}(s,t) $ the phase of which is independent on $ t $,which leads to the following 
relationship

\begin{equation}
 T_{N}(s;t)= T_{N} (s+i0,t)\vert _{s\in [s_{1},s_{2}]}= (1+i\omega(s))ReT_{N}(s;t).
 \end{equation}
Let us take use of the unitarity condition 
\begin{equation}
a_{l}(s) = a_{l}(s)^{2} + r_{l}(s)^{2} + \eta_{l}(s)
\end{equation}
 for partial waves where 
 \begin{equation}
 a_{l}(s) = \frac{1}{32\pi} \sqrt{1-\frac{4m^{2}}{s}} \int d\cos \theta P_{l}(\cos \theta)Im T_{N}(s,\cos \theta) ) 
 \end{equation}
 \begin{equation}
 r_{l}(s)= \frac{1}{32\pi} \sqrt{1-\frac{4m^{2}}{s}} \int d\cos \theta P_{l}(\cos \theta)Re T_{N}(s,\cos \theta)
 \end{equation}
 and $\eta_{l}(s)$ is the contribution of all inelastic channels,
 with
 $ t= -2p^{2}(1- cos \theta ) $.
 
 It follows from Eq.(3) that at $ s \in [s_{1},s_{2}]\subset [s_{inel}-\Delta s,s_{inel}) $  (when $ \eta_{l}(s) =0 $)
 we get the following "solution" of the unitarity equation ( $ a_{l}(s) = a_{l}^{2}(s) + r_{l}^{2}(s))$ :
 
  \begin{equation}
 r_{l}= \omega (s)/(1+\omega^{2}(s))= \rho(s)/(1+\rho^{2}(s)).
 \end{equation}
 From this we already see an evident disadvantage of this "solution": the partial wave amplitude does not decrease with the growth of angular momentum $ l $ as follows from the analyticity in $ t $ and polynomial boundedness in $ s $ \footnote{I am grateful to A. Samokhin for this observation.}  \cite{Ma}. 
 The very form of the "solution" (7) implies that
 \begin{equation}
 T_{N}(s;t)= \frac{\omega (s)(1+i\omega (s))}{(1+\omega^{2}(s))}\frac{16\pi\sqrt{s}}{\sqrt{s-4m^{2}}}\sum_{l\geq 0}(2l+1)P_{l}(\cos \theta)=
 \end{equation}
 \[= \frac{\omega (s)(1+i\omega (s))}{(1+\omega^{2}(s))}\frac{32\pi\sqrt{s}}{\sqrt{s-4m^{2}}}\delta(1-\cos \theta).\]
 Thus we see that at $ \theta \neq 0 $ and $s \in [s_{1},s_{2}]\subset [s_{inel}-\Delta s,s_{inel}) $  the scattering amplitude vanishes.
 
 It follows, in particular, that on the interval $s \in [s_{1},s_{2}]\subset [s_{inel}-\Delta s,s_{inel}) $  the discontinuity of the scattering amplitude is zero. Now, even with a modest assumption that the scattering amplitude is analytic in $ s $ in the vicinity of the s-channel physical region 
 $ [4m^{2},+ \infty) $ we conclude, again on the basis of the uniqueness property of analytical functions, that $ T_{N}(s,t) $ being vanishing on the segment $ [s_{1},s_{2}] $ should vanish in the whole analyticity domain, including the physical region.Q.E.D.
 \section{Conclusion and Discussion}
So, we see that - at first glance, not very restrictive - the condition of independence of the phase of the elastic scattering amplitude from the momentum transfer $ t $ in an arbitrarily small real neighborhood of the point $ t = 0 $ leads to a very significant result: the independence of the phase from momentum transfers \ textit {in the entire physical area} of the latter.

Accounting for the unitarity condition leads to the fact that in the region of purely elastic scattering (below inelastic thresholds) the phase of the amplitude along the momentum transfer \ textit {cannot be independent of $ t $} even on an arbitrarily small interval of transmitted momenta - otherwise the scattering amplitude turns out to be identically equal to zero.

The conclusion about the constancy of the phase as a function of the transmitted momenta given above does not use the fact that the real and imaginary parts of the scattering amplitude are not completely independent, because they are connected by dispersion relations due to analyticity  in energy . There are few strictly proved dispersion relations: they relate to pion-nucleon and kaon-nucleon scattering and for very limited momentum transfer values.

If you do not make too stringent requirements for rigor, then "at the level of heuristic considerations" and using dispersion relations, one can get the following result:

\textit {under the conditions of Theorem I and for "sufficiently high energies" the scattering amplitude is factorized:}

\begin {equation}
T_{N}(s;t)\mid_{s\rightarrow \infty} \Longrightarrow F(s))\cdot G (t).
\end {equation}

Such factorization is unacceptable, if only because it sharply contradicts the known experimental facts.
Thus, if we take into account the above as a sufficient basis, then the phase of the scattering amplitude \textit {cannot be independent of momentum transfers even at their arbitrarily small intervals.}

From the practical point of view  $ t $ -independence of the scattering phase can be implemented as "a weak $ t $-dependence "(at least at small $ t $) which can be implemented in the smallness of $ \frac{\partial\Phi_{N}}{\partial t} (s;0) $. At the  moment the only thing we know about the rate of $ t $ -dependence of the phase at small $ t $ is the relationship \cite{Kun} 
\begin{equation}
\frac{\partial\Phi_{N}}{\partial t} (s;0) = \frac{1}{4}tg(\Phi_{N}(s;0))[\langle b^{2}\rangle_{tot}- 2B(s)]
\end{equation}
where $ B(s) $ is the forward slope of $ d\sigma/dt $ (observed) and $ \langle b^{2}\rangle_{tot} $
is the total transverse extent of the interaction region (unobserved .
With $ t $-independent phase we would have 
$ \langle b^{2}\rangle_{tot} = 2B(s) $ (widely assumed).

 How weak such a rate can be without contradiction to basic principles is an open question at the moment.
 
 In Eq.(10) one can see the relationship of the scattering phase and its $ t $-dependence with the spatial characteristics, viz., \textit{the transverse range} $ \langle b^{2}\rangle_{tot} $ of the interaction region.
 
 There is a more direct relation if to consider also the \textit{longitudinal range.}
  The average longitudinal size of the interaction region in s.c.m.,  $ \langle \Delta x^{\ast}_{\parallel}\rangle $, is related to the phase of the scattering amplitude as follows \cite {Ptr}:
  \[\langle \Delta x^{\ast}_{\parallel}\rangle = 2\sqrt{s}\langle \frac{\partial \Phi_{N}(s,t)}{\partial t}\rangle = \frac{2\sqrt{s}}{\sigma_{el}}\int dt\frac{d\sigma_{el}}{dt}\frac{\partial \Phi_{N}(s,t)}{\partial t}. \] 
  
  If the phase is independent of $ t $, then 
  
  \[\langle \Delta x ^ {\ast}_{\parallel}\rangle = 0.\]

This, however, contradicts the long-known phenomenon of an increase in the longitudinal range of the interaction region with an increase of the collision energy \cite {Gri}.

 \section{Acknowledgements}
 I am indebted to Anatolii Samokhin for very fruitful discussions which helped to improve the formulation and  presentation of the results of this paper.
  
 This work is supported by the RFBR Grant 17-02-00120.


\begin{thebibliography}{99}

\bibitem{Ku}
V. Kundr\'{a}t and M. Lokaj\'{i}\v{c}ek,
 
Z. Phys. \textbf{C63}, 619(1994);

V. A. Petrov,

e-Print: arXiv:1906.03895 [hep-ph]
\bibitem{Be}
H. Bethe, Ann. Phys., \textbf{3} (1958) 190;

\bibitem{We}
G.B. West and D. R. Yennie, Phys. Rev. \textbf{172}, 1413(1968);

R. J. Glauber and G. Matthiae, Nucl.Phys.\textbf{B21} (1970) 135-157;

L. Baksay et al. Nucl.Phys. \textbf{B141} (1978) 1-28;

R. Cahn, Z. Phys. \textbf{C15}, 253 (1982);

G. Antchev et al. TOTEM Collaboration. 
Eur.Phys.J.\textbf{ C79 }(2019) no.9, 785;

J.R. Cudell, O.V. Selyugin, e-Print: arXiv:1901.05863 [hep-ph]

\bibitem{Ma}
Andr\'{e} Martin, 
Scattering Theory: Unitarity, Analyticity and Crossing. Lecture Notes in Physics. Springer Verlag, 1969.

\bibitem{Pri}
I. I. Privalov, 

Introduction to the Theory of Functions of a Complex Variable, GITTL, 

Moscow-Leningrad, 1948 (14n ed: 1999);

Henri Cartan,

Th\'{e}orie \'{e}l\'{e}mentaire des fonctions analytique d'une ou plusieurs 

variables complexes.

Hermann, Paris 1961;

E. C. Titchmarsh,

The theory of functions.

Oxford University Press ( Second edition, 1939).

\bibitem{Kun}
 	
Ji\v{r}\`{i} Proch\`{a}zka, Milo\v{s} V. Lokaj\'{i}\v{c}ek, Vojt\v{e}ch Kundr\'{a}t,

 Eur.Phys.J.Plus \textbf{131} (2016) no.5, 147
 
\bibitem{Ptr}
V. A. Petrov,


 EPJ Web Conf. \textbf{138} (2017) 02008.

\bibitem{Gri}
V. N. Gribov, B. L. Ioffe and I.Ya. Pomeranchuk,

Sov.J.Nucl.Phys.\textbf{ 2} (1966) 549, Yad.Fiz. \textbf{2} (1965) 768-776;

  B. L. Ioffe, 
  
  Phys. Lett.\textbf{30B}, 123 (1969);
  
  The latest discussion of the longitudinal extent of interactions can be found in
  
  G. A. Miller and S. J. Brodsky, 
  
  arXiv:1912.08911v3  [hep-ph]  3 Jan 2020.
  
\end{thebibliography}
\end{document}